\documentclass[a4paper,final]{aipproc}

\usepackage{epsfig,graphicx,epsf}
\usepackage{amssymb,amsfonts,amsmath}
\pdfoutput=1
\layoutstyle{8x11single}

\newcommand{\be}{\begin{equation}}
\newcommand{\ee}{\end{equation}}

\newcommand{\ba}{\begin{eqnarray}}
\newcommand{\ea}{\end{eqnarray}}

\begin{document}

\title{Searching for P- and CP-odd effects in heavy ion collisions}

\classification{}
\keywords      {}

\author{A.A. Andrianov}{
  address={V. A. Fock Department of Theoretical Physics, Saint-Petersburg State University,
198504 St. Petersburg, Russia}
}

\author{V.A. Andrianov}{
  address={V. A. Fock Department of Theoretical Physics, Saint-Petersburg State University,
198504 St. Petersburg, Russia}
}

\author{D. Espriu}{
  address={ICCUB, University of Barcelona, Mart\'\i ~i Franqu\`es, 1, 08928 Barcelona, Spain}
}

\author{X. Planells}{
  address={ICCUB, University of Barcelona, Mart\'\i ~i Franqu\`es, 1, 08928 Barcelona, Spain}
}

\begin{abstract}
In this talk we will summarize the main results from our recent work
concerning the possibility that a new metastable phase occurs in some
heavy ion collisions (HIC). This phase would be characterized by
the breaking of two characteristic symmetries of strong
interactions; namely $P$ and $CP$. We
investigate the experimental consequences of
parity breaking in such a situation and propose suitable
observables to elucidate the presence this phenomenon.
\end{abstract}

\maketitle

%%%%%%%%%%%%%%%%%%%%%%%%%%%%%%%%%%%%%%%%%%%%
%% MAINMATTER
%%%%%%%%%%%%%%%%%%%%%%%%%%%%%%%%%%%%%%%%%%%%

\section{Why P and CP might not be good symmetries in HIC}
Parity is one of the well established global symmetries of strong interactions.
While there are arguments to think that $P$ or $CP$ cannot be broken in
the usual QCD vacuum\cite{VW}, this may not be the case under extreme
conditions of temperature and density. Indeed no fundamental principle
forbids spontaneous parity breaking for $\mu\neq 0$ or out of equilibrium.

For a long time the possibility that
$P$- and $CP$-odd condensates may exist as regular phase at finite density
has been contemplated, going back all the way to the work
of Migdal\cite{mig}. The debate was inconclusive by
using simple nucleon-meson models and, if anything, the arguments suggested
that a condensate of this type was not physically viable. However, more recent
--and more complete--  effective theory studies
indicate that such a phase
is a real possibility\cite{anesp} at `moderate' densities
(3 to 8 $\times\, \rho_N$) compared to the normal nuclear density $\rho_N ~ 0.17 fm^{-3}$.

This possibility is surely relevant in astrophysical contexts
but it may still be posible to ephemerally produce
such a phase in HIC in conditions where the density is high (i.e.
large baryonic chemical potential) and temperature relatively low.
These conditions may be within reach of future experiments at FAIR
and NICA\cite{nica}. However this is not the main subject
of this presentation.

Rather we would like to consider the possibility that
in a violent collision 
long-lived bubbles with a non-vanishing value of the topological charge,
where parity is locally broken, could be produced. The possibility that these
 topological fluctuations
would take place in a finite volume and  large $T$ was first
proposed by Kharzeev, Pisarski and Tytgat\cite{kpt} and
later considered in a number of works\cite{other,aaep}

Large fluctuations in the gauge field could indeed exist in a hot
environment and they could generate
a local imbalance of topological charge. This picture is supported
by lattice simulations\cite{lattice} although
results obtained in Euclidean simulations are hard to connect
with the time dependent dynamics that
exist in the early stages of a HIC.

The possibility of domains where $P$ is broken is also supported by the
{\em glasma} picture\cite{glasma}.
However domains in this model are typically very small ($<<1$ fm)
in the initial QGP phase. Of course these domains should expand and grow
at later stages.  Actually in our work we will be interested exclusively in
the hadronic phase that develops later in the collision.
We will assume that some domains with a net topological
charge and spatial extent $> 1$ fm exist in this hadronic phase,
originally created via the glasma mechanism or any other.

\subsection{Generating an effective chiral chemical potential}
If a topological charge $T_5$
\be
T_5=\frac{1}{8\pi^2}\int_{\text{vol.}}d^3x\varepsilon_{jkl}\text{Tr}
\left (G^j\partial^k G^l-i\frac23G^jG^kG^l\right )
\ee
arises in a finite volume due to quantum fluctuations of gluon fields and it is
conserved for a sufficiently long time in a quasi-equilibrium
situation, we can
introduce a topological chemical potential $\mu_\theta$ conjugate to it
into the QCD lagrangian via $\Delta \mathcal L_{top} =\mu_\theta \Delta T_5$
that plays the role of an effective $\theta$ term. The variation of
the topological charge is gauge invariant
\be
\Delta T_5 = T_5(t_f) - T_5(t_i) = \frac{1}{8\pi^2}\int_{t_i}^{t_f} dt
\int_{\text{vol.}}d^3x \text{Tr}\left (G^{\mu\nu} \widetilde G_{\mu\nu}\right ).
\ee

We will assume that, temporarily, as a consequence of
a topological fluctuation of gluon fields or some other mechanism an effective $CP$- odd
large $\theta$ term is generated. We will
suppose that this region eventually grows in the hadronic phase
to a sufficiently large size. We will also assume
that this situation can be treated by equilibrium
field theoretical methods.

For {\em peripheral} heavy ion collisions such
a situation may lead to the Chiral Magnetic Effect\cite{cme} whereby
a large $\theta$ term, combined with the
large magnetic field due to the colliding nuclei, generates a large
electric field and originates charge separation.

For {\em central} collisions (and light quarks) a large $\theta$ term
will trigger an ephemeral phase with axial chemical potential $\mu_5 \neq 0$.
This comes about because the PCAC equation
predicts an induced axial charge to be conserved in the chiral ($m=0$) limit:
\be
\frac d{dt}\left (Q_5^q-2N_f T_5\right )\simeq 0,
\quad Q_5^q=\int_{\text{vol.}}d^3x\bar q\gamma_0\gamma_5q = \langle N_L - N_R\rangle,
\ee
the latter being an average chiral state asymmetry.
Neglecting the quark mass is a good approximation for the lightest
$u$ and $d$  quarks only. The strange quark is already too heavy and erases
the chiral charge in the time scales where this phenomenon could be
relevant. In a quasi-equilibrium situation the appearence of a nearly
conserved chiral charge can be incorporated with the help of a
chiral chemical potential $\mu_5$.

\section{Effective meson theory in a medium with local parity breaking}
In principle, from the discussion in the previous section we have two possible isospin
structures for $\mu_5$: (a) An isosinglet pseudoscalar background is
expected to be relevant in situations where temperature is the main external driver ($T\gg\mu$)
such as in RHIC and LHC, and it would be due to the formation of domains with
a non-zero topological charge.
(b) An isotriplet pseudoscalar background could be appropriate in situations where $\mu\gg T$,
as will be the case in FAIR or NICA, and a true thermodynamic phase
forms temporarily. The formation of an isosinglet pseudoscalar condensate cannot be excluded
in this case either.
Only the situation (a) will be considered in this talk.

Having in mind that although the parity breaking domain may have formed in the early
stages of the HIC we are only concerned about its consequences in the later stage
hadronic phase we will proceed to build an effective theory for mesons in a
$P$- and $CP$-odd environment.
We will deal in turn with scalars and vector mesons.
Note that in both cases a breaking of Lorentz symmetry will occur.

The $J=0$ sector can be described by using the spurion technique in the Lagrangian
\be
D_\mu \Longrightarrow  D_\mu - i \{\mu_5 \delta_{0\mu}, \cdot \}
\ee
and constructing a generalised sigma model\cite{dalitz}
 with the light scalar mesons $\sigma,\vec \pi,\eta,\eta',\vec a_0$. The
most relevant operator has dimension $D=2$. All these hadronic states naturally have a well defined parity
in `normal' QCD and partly because of that only the $\sigma$ and $\rho$ interact strongly with
the pion fireball.

The new eigenstates of the hamiltonian do not have a well defined parity.
Now due to parity breaking there is mixing in the $\sigma-\eta-\eta'$
and $\vec \pi-\vec a_0$ channels. The resulting $J=0$ eigenstates are {\em all}
coupled and are expected to {\em thermalize} in the HIC fireball.

For vector mesons the most relevant operator has the dimension $D=3$.
$P$- and $CP$-odd  effects will appear through the Chern-Simons term\cite{aaep,prl}
\be
\Delta\mathcal L\simeq \varepsilon^{\mu\nu\rho\sigma}\text{Tr}\left [\hat\zeta_\mu V_\nu V_{\rho\sigma}\right ]
\ee
with $\hat\zeta_\mu= \partial_\mu \hat a(\vec x,t)= \delta_{\mu 0} \partial_0 \hat a(t)$
for a spatially homogeneous, time dependent background field $\hat a(\vec x,t)$.
We shall assume here $\partial_0 \hat a(t)\sim \hat\zeta\propto \mu_5$, as follows from the discussion
in the previous section.

Vector mesons will be introduced and treated in the conventional way using the
Vector Meson Dominance model\cite{vdm} and enter the above lagrangian via the corresponding
vector current $V_\mu$. Note that for vector mesons there will be no mixing at this order
resulting from the previous lagrangian but rather a distortion of the spectrum.

The current $V_\mu$ formed by a combination of vector mesons and the photon
couples to fermions
\be
{\cal L}_{\text{int}} = \bar q \gamma_\mu \hat V^\mu q;\quad \hat V_\mu \equiv -
e A_\mu Q  + \frac12 g_\omega  \omega_\mu \mathbb{I} + \frac12g_\rho \rho_\mu^0  \tau_3,
\ee
where  the charge $Q= \frac{\tau_3}{2} + \frac16$ and the quark-meson coupling constants $ g_\omega \simeq  g_\rho \equiv g \simeq 6 $.
In addition we have to include Maxwell and mass terms
\be
{\cal L}_{\text{kin}} = - \frac14 \left(F_{\mu\nu}F^{\mu\nu}+
\omega_{\mu\nu}\omega^{\mu\nu}+ \rho_{\mu\nu}\rho^{\mu\nu}\right)
+ \frac12  V_{\mu,a}(\hat m^2)_{a,b}V^\mu_{b}
\ee
\be
\hat m^2\simeq m_V^2\left(\begin{array}{ccccc}
\frac{10 e^2}{9g^2} & &-\frac{e}{3g} && -\frac{e}{g}\\
 -\frac{e}{3g}&& 1 && 0\\
 -\frac{e}{g} && 0 && 1\\
\end{array}\right), \quad \mbox{\rm det}\left(\hat m^2 \right) = 0.
\ee

Particularizing to the present situation we get the following
$P$-odd interaction
\be\label{poddint}
{\cal L}_{P-odd}(k)= \frac12 \zeta \epsilon_{jkl}\, V_{j,a} \,N_{ab}\,\partial_k V_{l,b},
\ee
where in the case of an isosinglet pseudoscalar background
\be
N_{ab}^\theta\, \simeq\,  \left(\begin{array}{ccccc}
\frac{10 e^2}{9g^2} & &-\frac{e}{3g} && -\frac{e}{g} \\
 -\frac{e}{3g}&& 1 && 0 \\
 -\frac{e}{g} && 0 && 1 \\
\end{array}\right),\quad \mbox{\rm det}\left(N^\theta\right) = 0.
\ee
Simultaneous diagonalization of the matrices $\hat m^2, N$ leads to
\be
N = \mbox{\rm diag}\left[0,\, 1,\, 1+ \frac{10 e^2}{9g^2}\right]
\simeq \mbox{\rm diag} \left[0,\, 1 ,\, 1\right]
\quad
\hat m^2  = m_V^2 \, \mbox{\rm diag} \left[0,\, 1,\, 1+ \frac{10 e^2}{9g^2}\right]
\simeq \mbox{\rm diag} \left[0,\, 1,\, 1\right].
\ee
Note that after diagonalization the photon itself is unaffected as it
remains massless.

Due to (\ref{poddint}) different vector meson helicities get a different
momentum-dependent correction to its mass\cite{AS}.
Vector mesons exhibit the following dispersion relation:
\be
m_{V,\epsilon}^2=m_V^2-\epsilon\zeta|\vec k|,
\ee
where $\epsilon=0,\pm 1$ is the meson polarization.
The position of the poles for $\pm$ polarized mesons is moving with wave vector $|\vec k|$.
Massive vector mesons split into three polarizations with masses $m^2_{V,+} < m^2_{V,L}< m^2_{V,-}$.
This splitting unambiguously signifies $P$ breaking. Could the splitting be measured?

\section{Possible manifestations of P-odd effects in HIC}
When trying to understand the nature of the fireball produced in a HIC it is
natural to investigate electromagnetic probes such as photons and leptons. In
the previous section we have shown that $P$ breaking has substantial effects on
the meson spectrum and this could eventually reflect in their leptonic decay products.

Let us then proceed to investigate possible `anomalies' in dilepton production
in various meson decays such as $\rho,\omega\to e^+ e^-$.
The total dilepton production also receives potential contribution from
the pseudoscalar Dalitz decays
$\eta,\eta'\rightarrow \gamma e^+e^-$
and the $\omega$ Dalitz decay $\omega\rightarrow \pi^0 e^+e^-$

In fact dilepton production shows a number of anomalies \cite{phenix,star,ceres}. Perhaps the most
obvious one is a surprising enhancement in the low dilepton invariant mass
region that has been observed in virtually all experiments for a long time.
This excess is to this date unexplained.

We show below data from the two RHIC experiments: PHENIX\cite{phenix} and
STAR\cite{star} at BNL. Other experiments have also seen a similar excess\cite{ceres}.
It is impossible to explain this excess with the standard `hadronic cocktail'.

\begin{figure}[ht]
\centering
\includegraphics[scale=0.2]{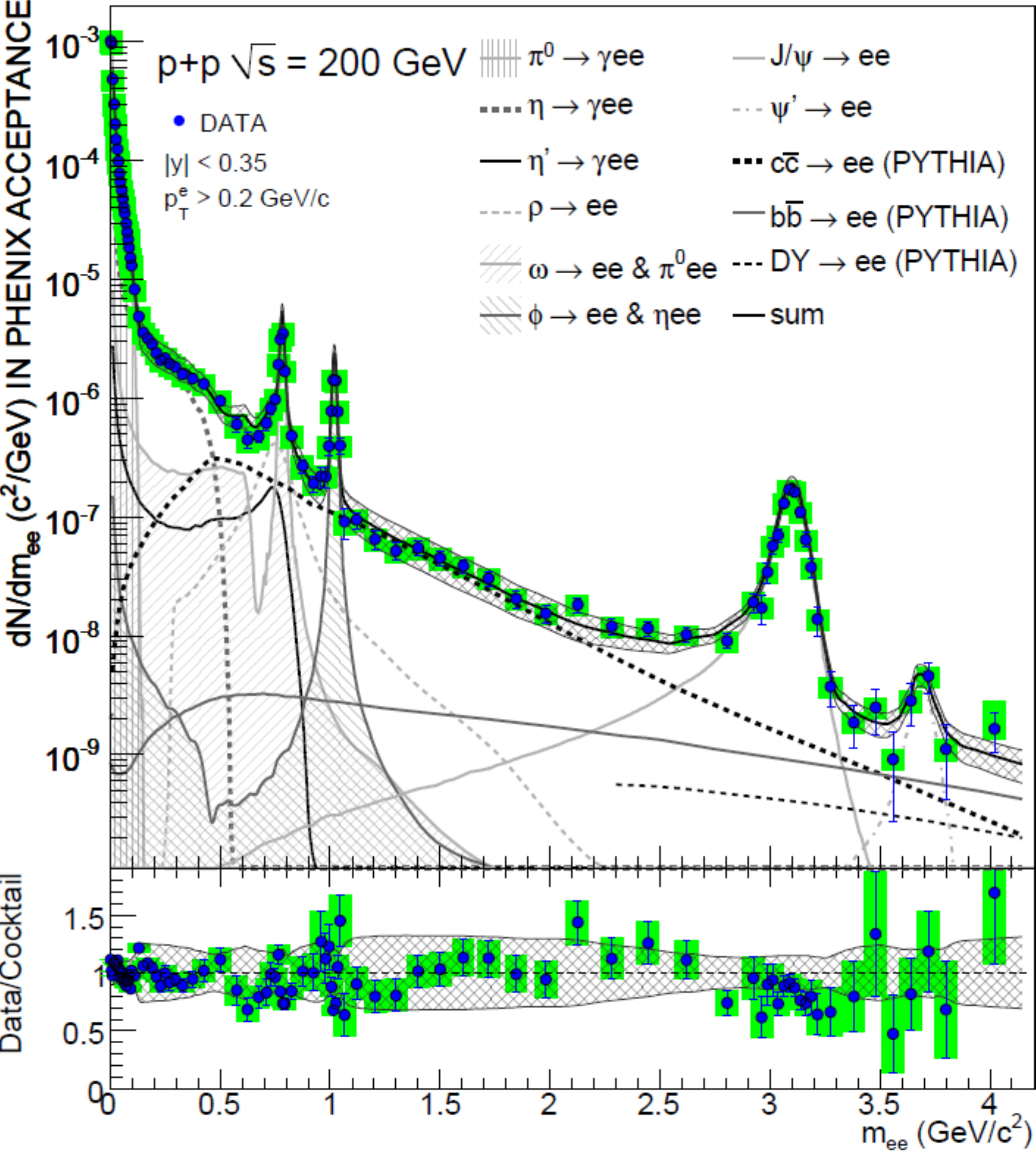} \quad \includegraphics[scale=0.21]{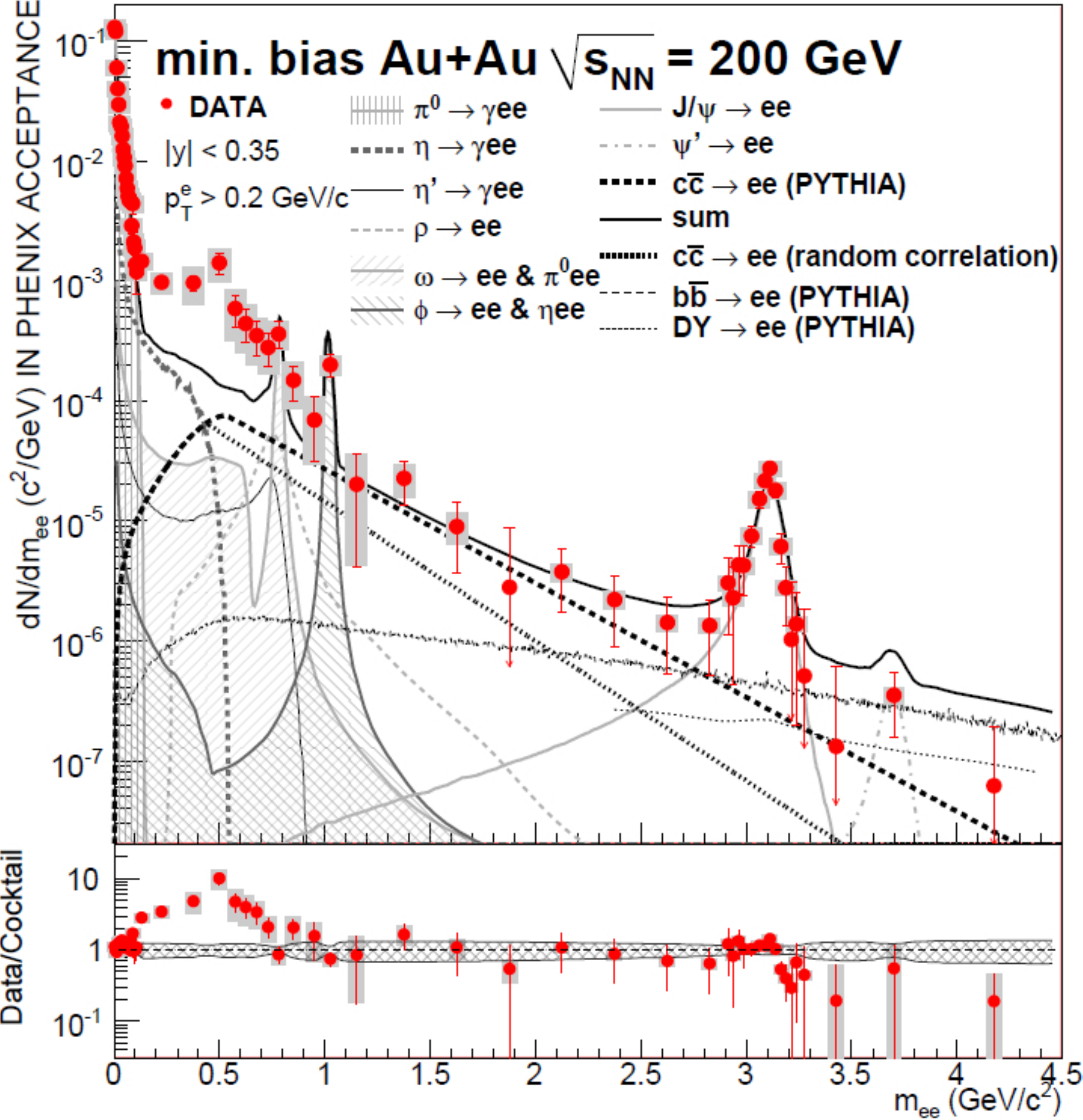}
\caption{Left: the yield of dileptons ($e^+e^-$ pairs) is exceedingly well reproduced in
proton-proton collisions by the hadronic cocktail. Right: however for {\em central} or {\em quasicentral}
collisions the cocktail fails completely to reproduce the data below the $\phi$ meson resonance. Figures
from the PHENIX collaboration.}
\end{figure}
\begin{figure}[ht]
\centering
\includegraphics[scale=0.4]{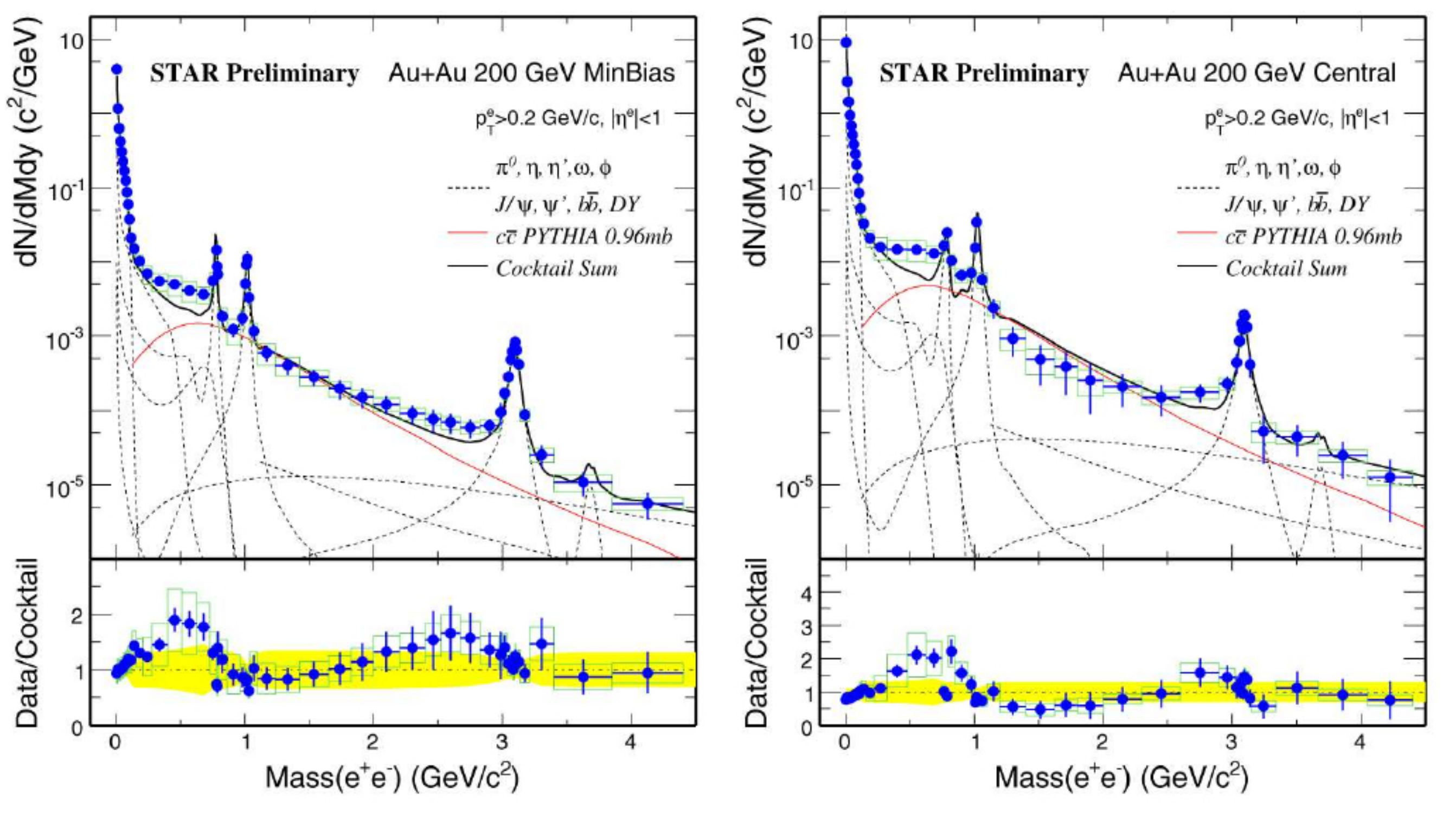}
\caption{Similar results as in PHENIX are obtained by the STAR collaboration. On the left the enhancement is
show for minimum bias events, whereas the right figure only contains central events. This shows that the effect
is really present for central or nearly central events. The enhancement is in any case less spectacular
than in PHENIX.}
\end{figure}
Another peculiarity is a
 large broadening of the $\rho$ spectrum. This was very clearly observed by the
NA60 collaboration\cite{NA60}
several years ago by
measuring dimuon production around the combined $\rho-\omega$ peak and carefully subtracting
contributions from Dalitz decays. Interpreting these results is not easy and it has remained
a big puzzle for a long time. On the other hand, the results seem to give credence to the idea of
`broadening' for in-medium $\rho$ mesons as opposed to the `shifting mass' scenario.
It is claimed that the large broadening can be understood by 'conventional' mechanisms although
this involves a certain amount of parameter fitting and it is certainly more of a post-diction
than a prediction. It is natural to question whether this issue is really understood.
\begin{figure}[ht]
\centering
\includegraphics[scale=0.40]{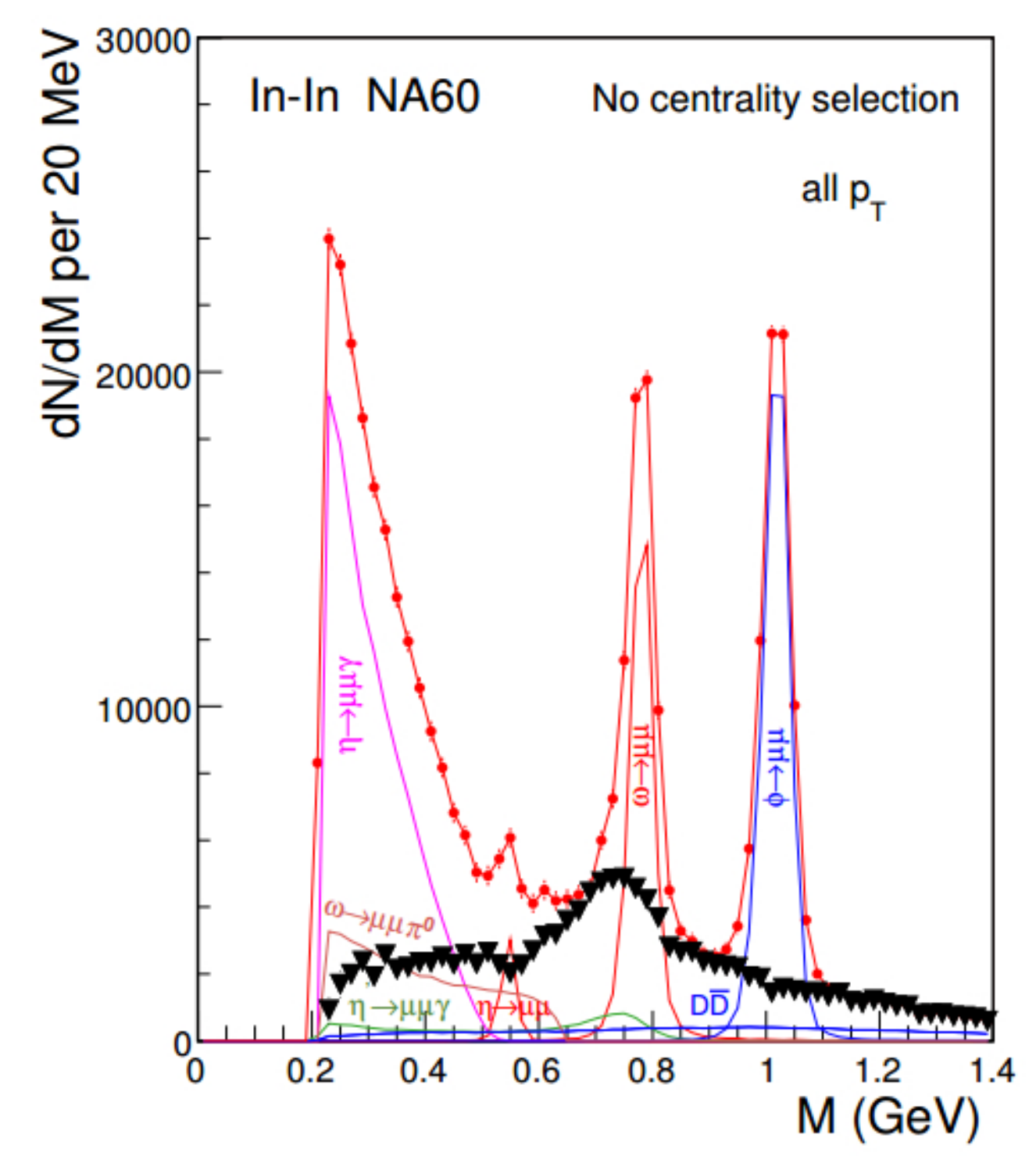}
\caption{A large broadening of the $\rho$ spectral function was measured in a neat way by the NA60 collaboration.}
\end{figure}

\subsection{ Dalitz decays in a P-odd environment}
Having seen that there a number of features in dileptons produced in HIC it is
legitimate to ask oneself if local parity breaking could be of some
relevance to account for, perhaps in combination with more conventional explanations,
some of these effects.

Let us first turn to the modifications induced in the $J=0$ sector, which contributes
to dilepton production via Dalitz decays. In what follows we will always
compare to the PHENIX measurements and for that reason we will use the same set of experimental cuts:
$|y_{ee}|<0.35$, $|\vec p_t^e|>200$ MeV and gaussian $M_{ee}$ smearing (width=10 MeV),
and the effective temperature $T=220$ MeV.

\begin{figure}[ht]
\label{leptondalitz}
\centering
\includegraphics[scale=0.12]{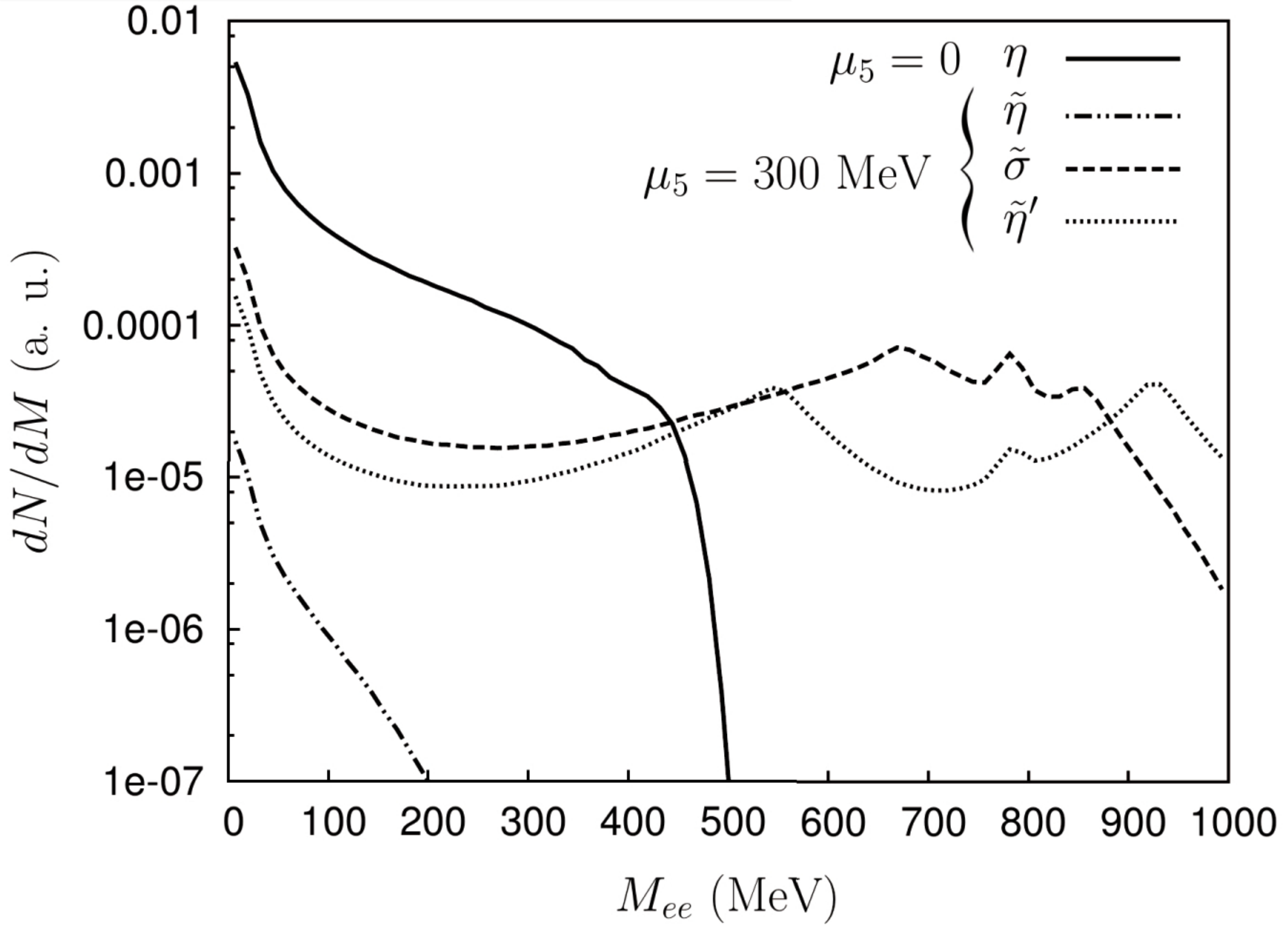}\includegraphics[scale=0.12]{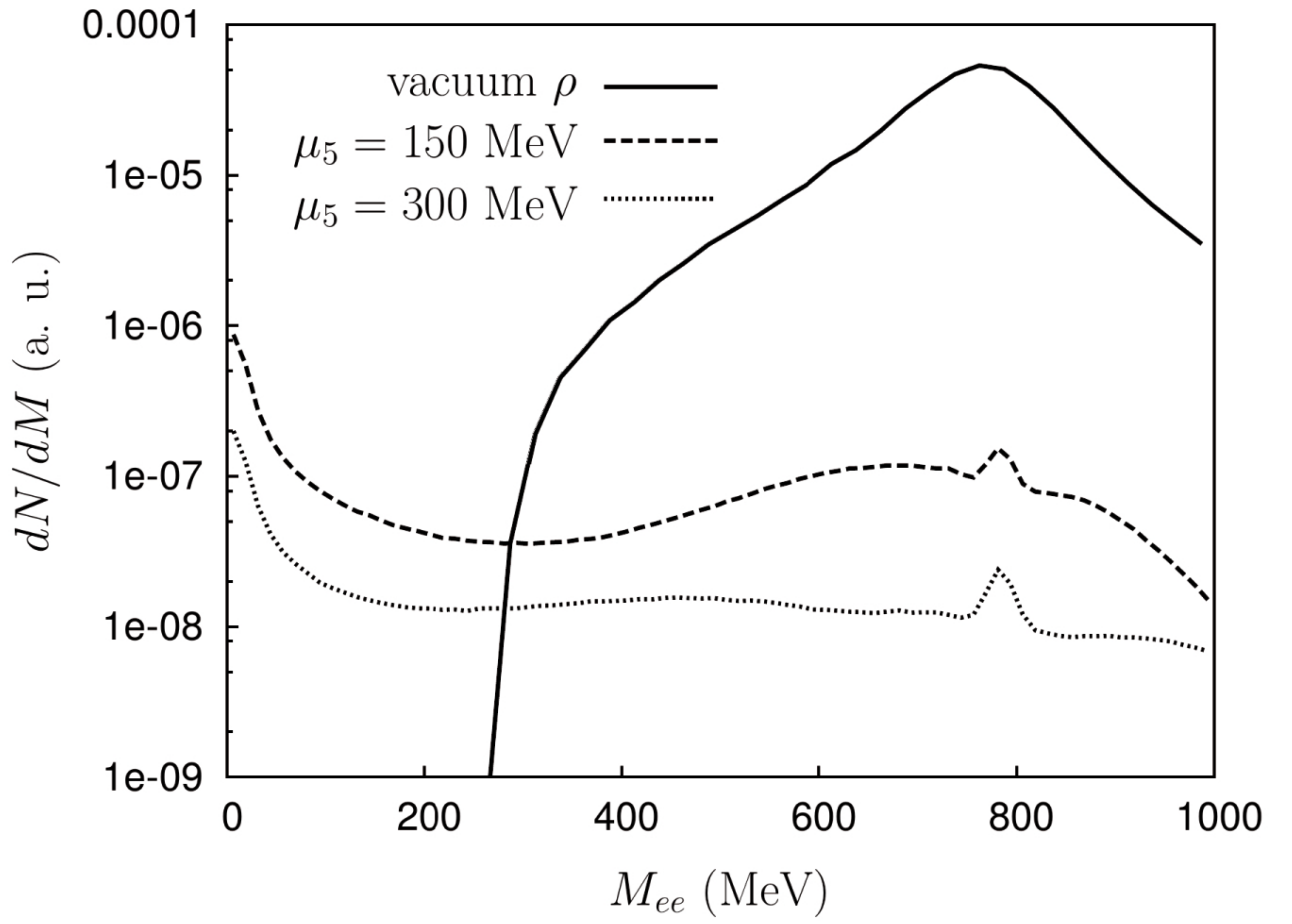}
\caption{Left: the production from the three new eigenstates is compared to the one from $\eta$ decays (that is the
only relevant one if undistorted vacuum properties are assumed). The latter is depleted but there is a net
enhancement at larger invariant masses. Right: the total dilepton production from the discussed mechanism
is shown for two values of $\mu_5$ and compared to the vacuum $\rho$ peak as a refernce.
However be warned that beyond $\mu_5\sim 100$ MeV the effective
lagrangian technique shows clear signs of failure.}
\end{figure}

Using the effective lagrangian briefly described before for $J=0$ mesons
we determine\cite{dalitz}
the spectrum (that as described contains states without definite parity) and, given
that they mix among themselves strongly, assume that they thermalize in the fireball
that now consists mostly of the lightest of such states. After that we proceed to compute
the corresponding Dalitz decays. Many more details can be found in X.Planells'
Ph D thesis\cite{thesis}. The corresponding predictions for $e^+e^-$ production
are shown in Fig. \ref{leptondalitz}

The results represent a net enhancement to dilepton production. Even though they are
clearly insufficient to explain the dramatic enhancement in PHENIX in the region 200-500 MeV, they help
vis a vis STAR data. However, before jumping to conclusions
the reader should be warned that except for very low
values of $\mu_5$ the effective lagrangian so obtained ceases to make sense very soon, a
fact that may reflect the need to introduce more resonances. Thus this approach is
of limited validity and we do not exclude that a more accurate treatment may actually provide
a higher dilepton yield and be part of the solution of the dilepton puzzle in the
200-500 MeV region. We do regard this as a totally open issue at present because unfortunately 
we have little phenomenological intuition of the properties of `broken parity' QCD.

\subsection{$\rho$ meson broadening}
\begin{figure}[ht]
\label{ourbroadening}
\centering
\includegraphics[scale=0.30]{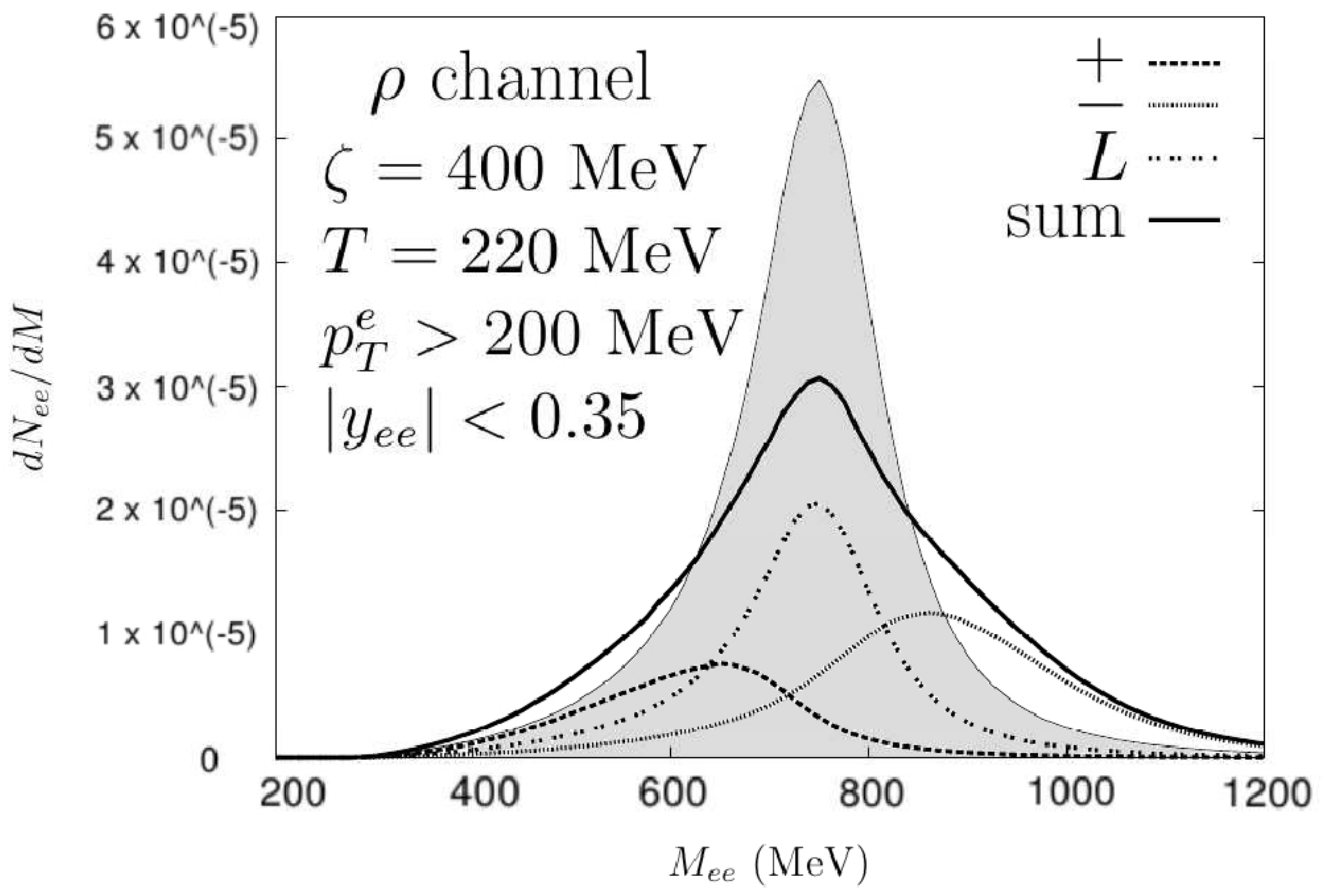}
\caption{Polarization splitting in the $\rho$ spectral function with local 
parity breaking  $\zeta=400$ MeV ($\mu_5=290$ MeV) compared
with $\zeta=0$ (shaded region). Note that the temperature is an effective or `boosted' one so it is not surprising
that it is actually larger than the deconfining temperature.}
\end{figure}

$P$-odd effects introduce broadening in a totally natural manner. Because masses depend
on the polarization, the original Breit-Wigner actually splits in three different peaks. This is clearly shown
in Fig. \ref{ourbroadening}
There is a similar effect for the $\omega$ meson.

It is of course very tempting to compare this `automatic' broadening with the very precise
experimental results
obtained by the NA60 collaboration from dimuon production. And indeed
the peculiar shape of the $\rho$ spectrum measured by  NA60 is grossly reproduced with
amazingly little effort. We used
our best guess from the available published data by the NA60 collaboration itself to implement the
experimental cuts. As the reader can see, we have not attempted to superimpose the two plots
beacuse the NA60 data is not properly acceptance corrected. We have tried hard to obtain
more information to make a meaningful comparison but found the collaboration
to be unresponsive.

In any case the comparison is tantalizing. Within the assumption of local parity breaking only one
parameter is fitted ---the value of $\mu_5$; the best fit is obtained for values of $\mu_5$ close
to the constituent mass. Incidentally these values are preferred in NJL type analysis for a stable
parity breaking phase to exist\cite{NJL}
\begin{figure}[ht]
\centering
\includegraphics[scale=0.85]{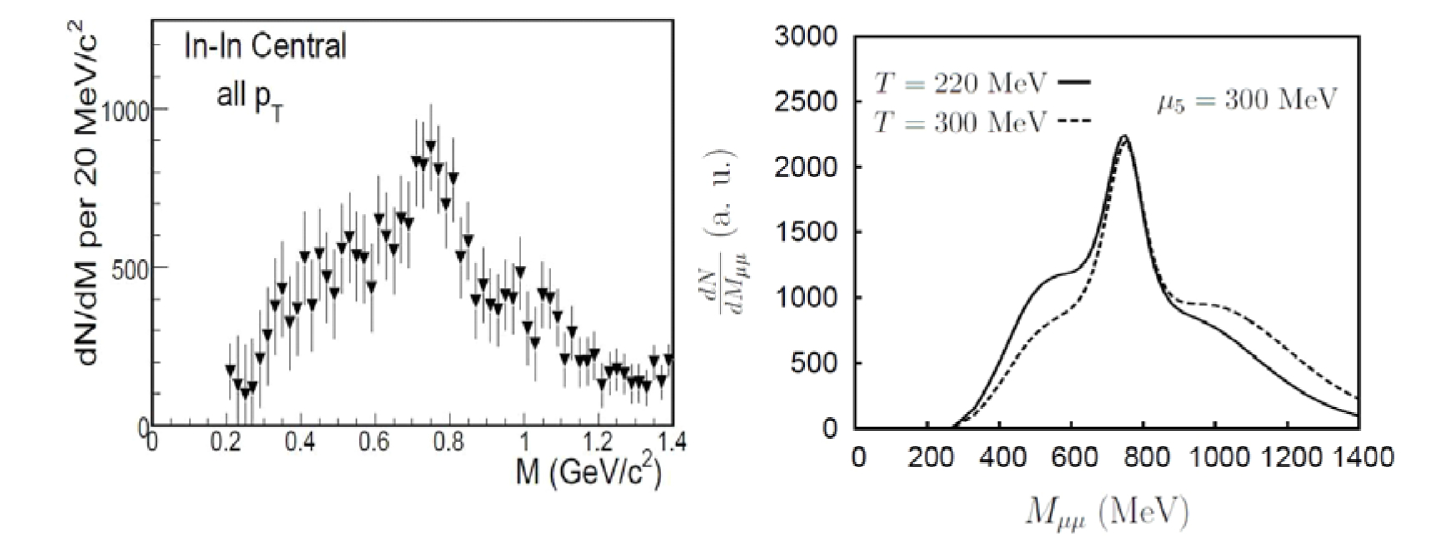}
\caption{Left: NA60 results for the $\rho$ spectral line-shape as a funtion of the dimuon invariant
mass. Right: the analogous quantity obtained from assuming local parity breaking with a value of
$\mu_5=300$ and two values for the effective temperature}
\end{figure}

\subsection{In-medium vector meson decays $V\to \ell^+\ell^-$}
\begin{figure}[ht]
\label{rhoomega}
\includegraphics[scale=0.30]{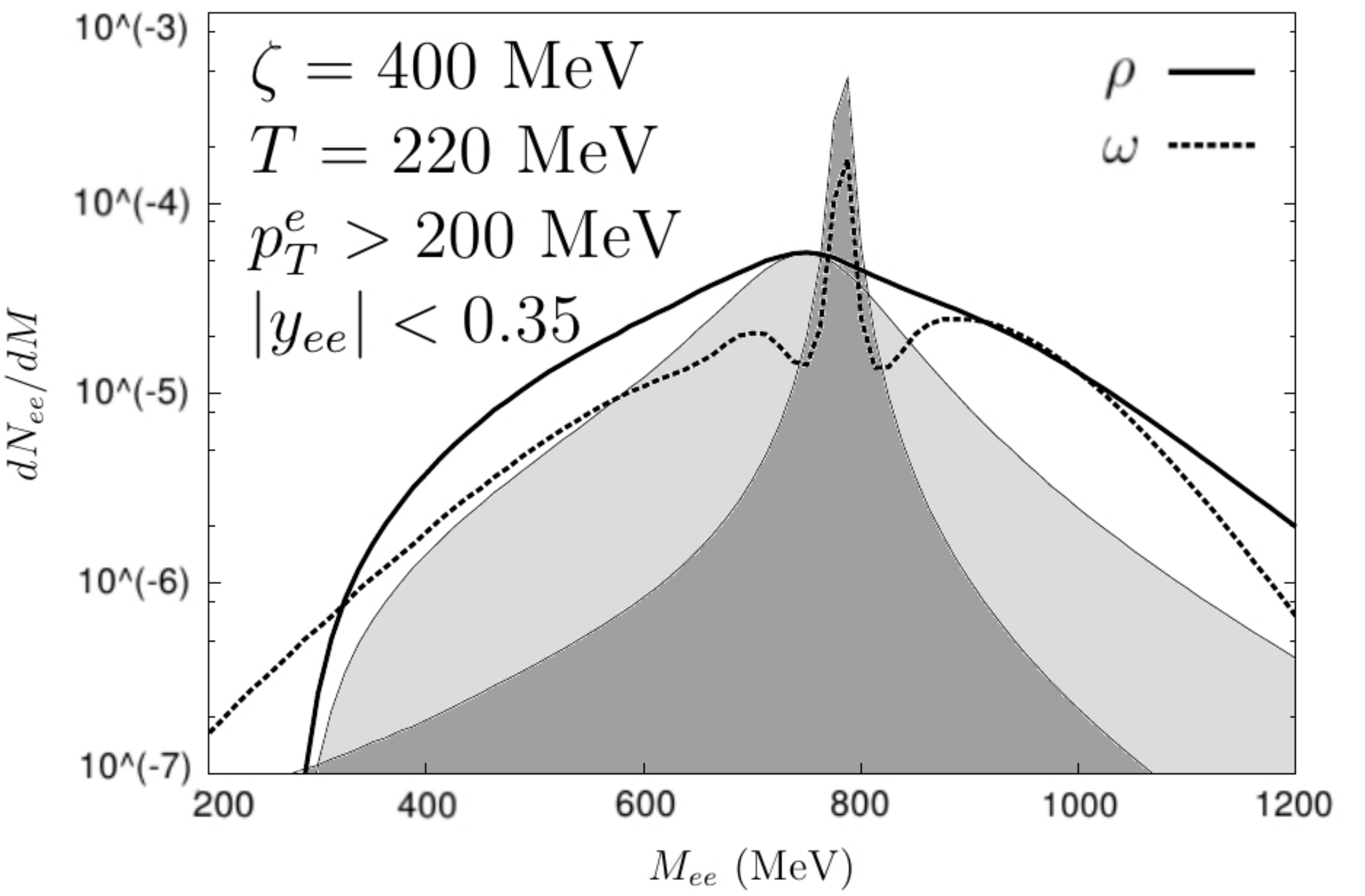}
\caption{In-medium $\rho$ and $\omega$ channels (solid and dashed line) and their vacuum contributions
(light and dark shaded regions) for $\mu_5=290$ MeV.}
\end{figure}

We shall be even briefer here, as this particular point has already been reported and discussed
in several conferences. We will now include both the $\rho$ and $\omega$ mesons and consider
the corresponding distortions in their respective spectra for dilepton production. The results
are shown in Fig. \ref{rhoomega}.

The associated enhancement of the dilepton yield  could (at least partly) explain the
anomalous enhancement seen by PHENIX and STAR.

\section{Observables sensitive to P-odd effects}
One of the unambiguous signals of $P$-odd effects is the separation between polarizations.
Is there any way to study these decays in order to separate the different polarizations
and thus confirm or rule out the presence of local parity breaking in HIC?

It is well known that the angular distribution of leptons carries information on
the polarizatio of the decaying mesonn. However, current angular distribution studies are not thought to
detect possible $P$-odd effects.

\begin{figure}[ht]
\label{angles}
\includegraphics[scale=0.4]{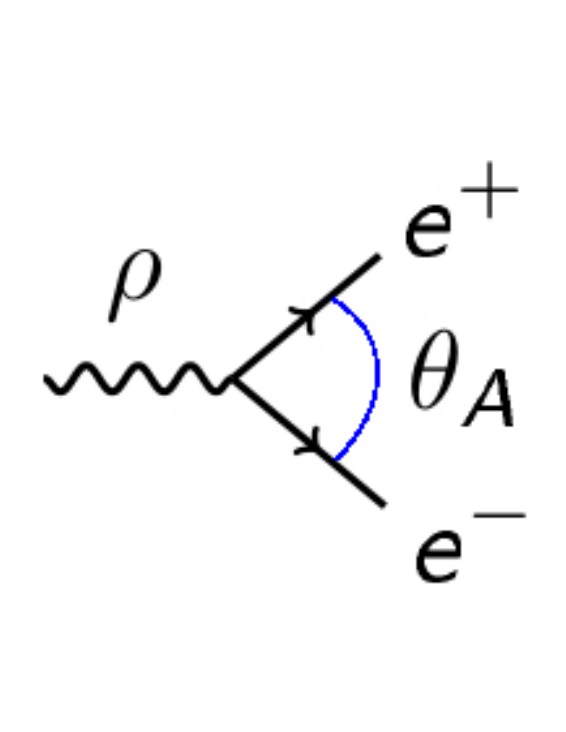}\qquad \includegraphics[scale=0.4]{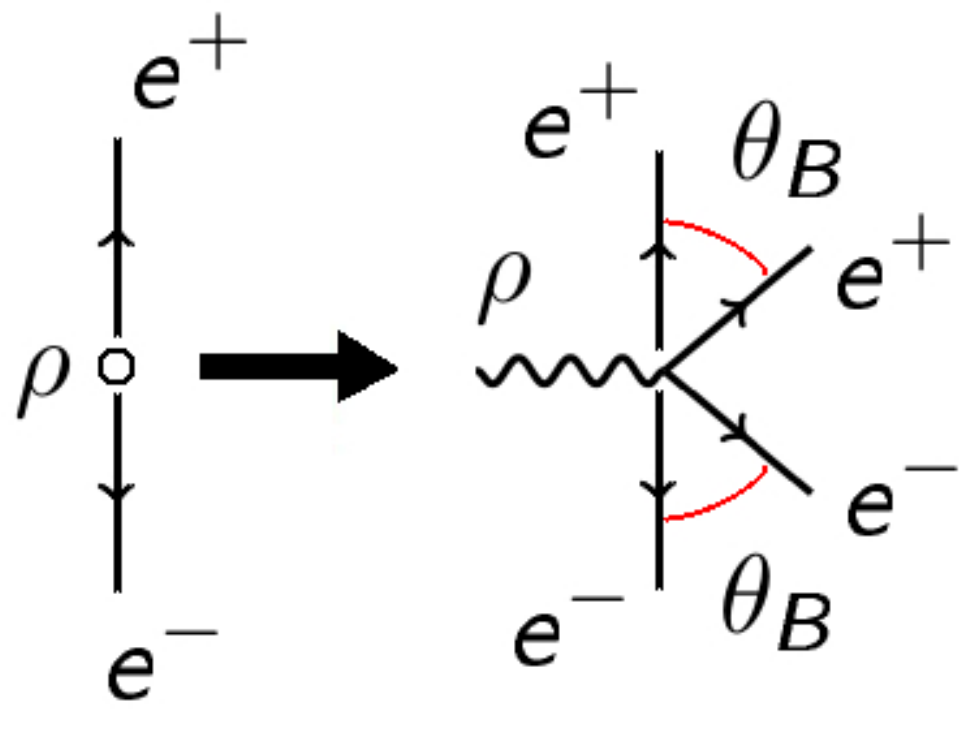}
\caption{Case A: $\theta_A$ is the angle between the two outgoing leptons in the laboratory frame.
Case B: $\theta_B$ is the angle between one of the two
outgoing leptons in the laboratory frame and the same lepton in the dilepton rest frame. }
\end{figure}

We will instead define two angles as described in Fig. \ref{angles}.
In order to isolate the transverse polarizations, we will perform different cuts for each angle and study
the variations of the $\rho$ (and $\omega$) spectral function. The results are given in the next
figures.

\begin{figure}[ht]
\centering
\includegraphics[scale=0.35]{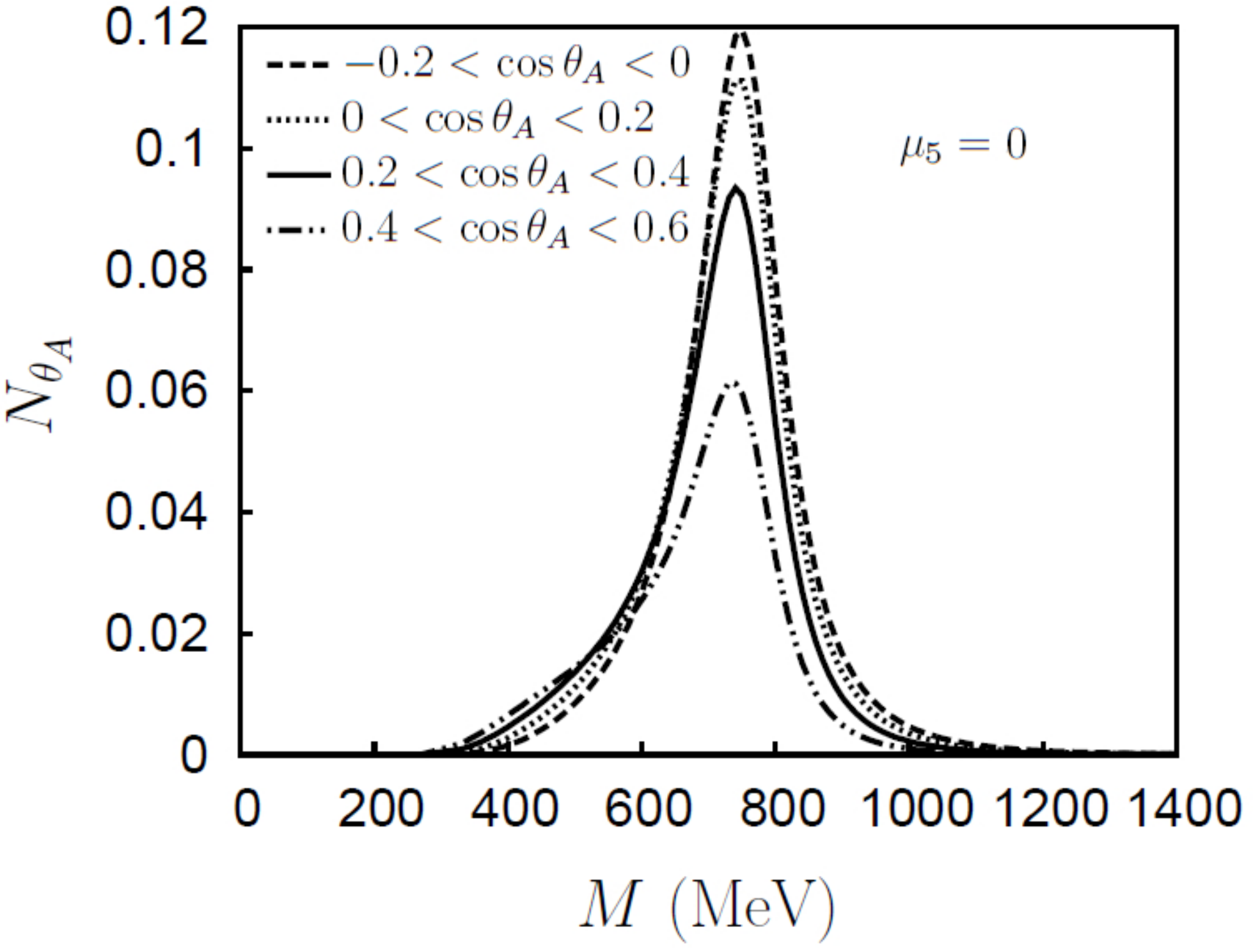}\includegraphics[scale=0.35]{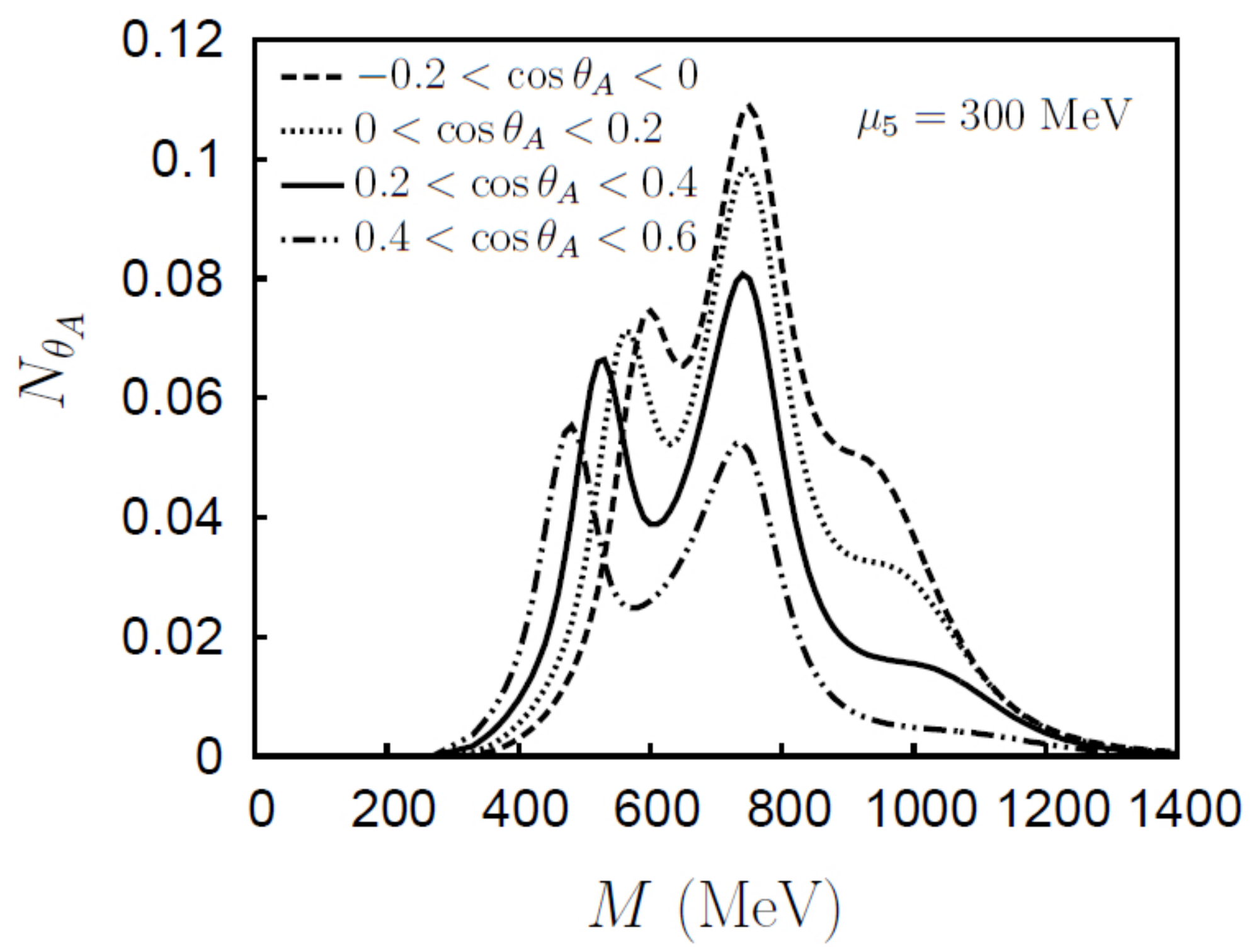}
\caption{Angle $\theta_A$ between the two outgoing leptons in the laboratory frame. $\rho$
spectral function depending on the dielectron invariant mass $M$ in vacuum ($\mu_5=0$)
and in a parity-breaking medium with $\mu_5=300$ MeV for different ranges of $\theta_A$.}
\end{figure}

A quite visible secondary peak appears in a $P$-odd medium! Note however that due to the cuts (needed to
make the secondary peak visible) there is an important reduction of the number of events:
the vacuum peak shows at
 most about $ 10\%$ of the events one would expect without any cut in $\theta_A$.

If the secondary peak is found for a particular angular coverage, its position would be an
unambiguous measurement of  the effective or mean value for $\mu_5$. This value of course need not be
the same for each HIC. It need not even be uniform. One only needs that domains are sufficiently large
for the $\rho$ meson to decay in the medium. The vacuum $\rho$ peak hides the secondary
one for $\mu_5\simeq 100$ MeV or below due to its
large width. For $\omega$, all the peaks are visible.

\begin{figure}[ht]
\centering
\includegraphics[scale=0.35]{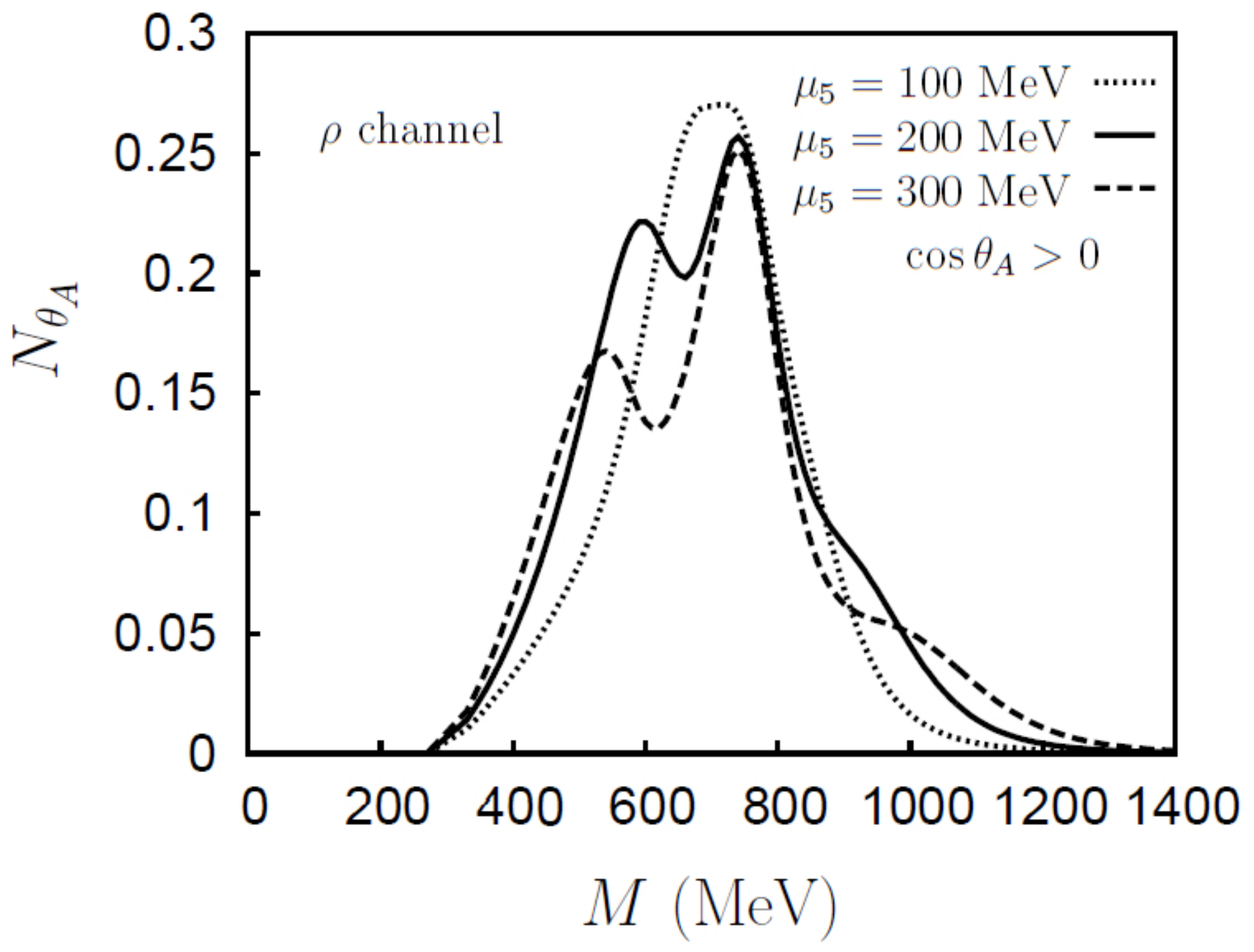} \includegraphics[scale=0.35]{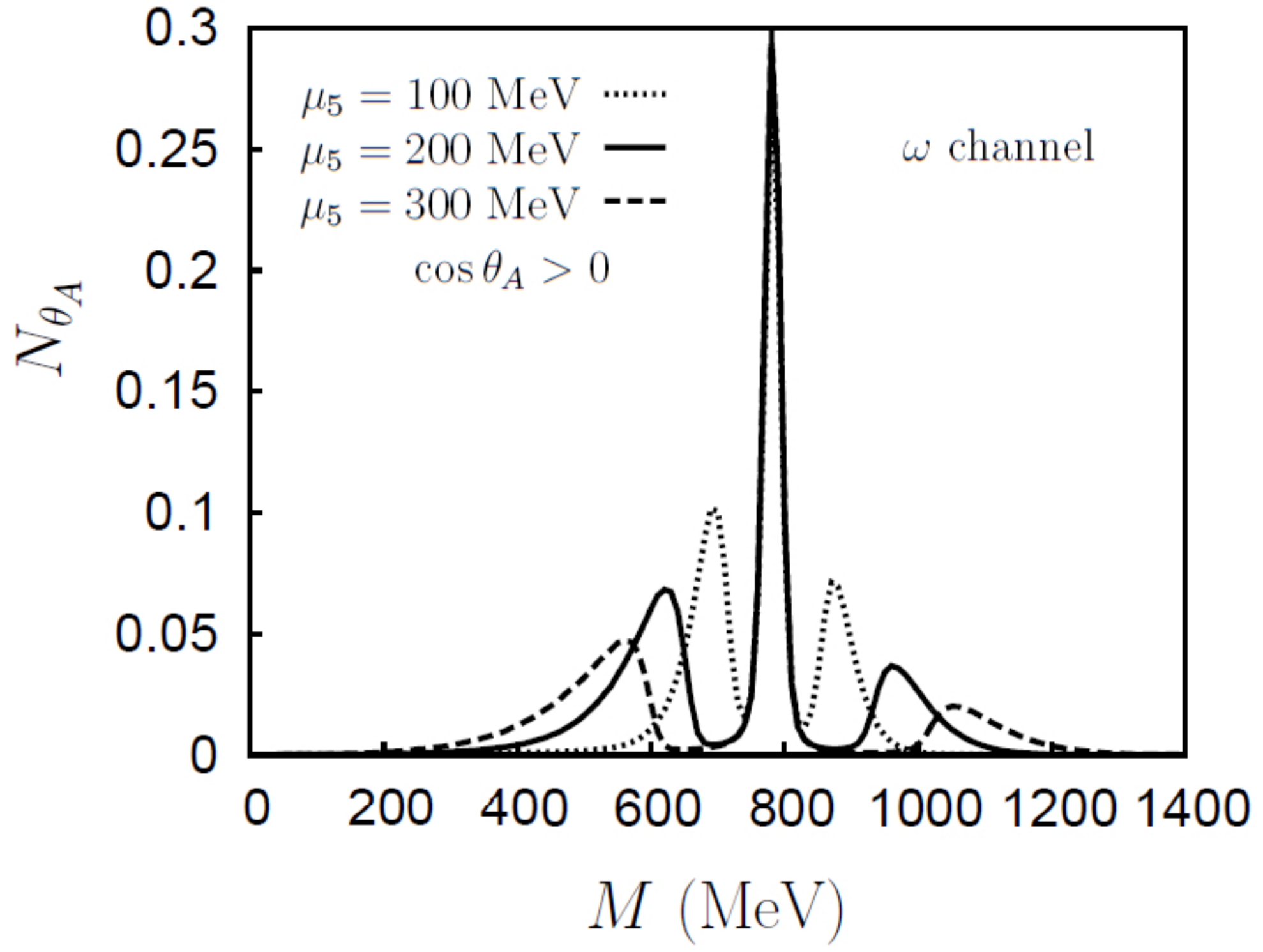}
\caption{$\rho$ and $\omega$ spectral functions depending on the invariant mass $M$ and
integrating $\cos\theta_A\geq 0$ for $\mu_5=100,200$ and 300 MeV.}
\end{figure}

We also present the combination of the $\rho$ and $\omega$ channels. In this case, the total production
is normalized to PHENIX data. It may seem questionable to assume that the $\omega$ decay inside
the firewall (and this is one reason to carefully separate both contributions). However, local parity
breaking should increase substantially the number of omega mesons decaying. For one thing there are
the by now familiar parity considerations; parity is no longer a conserved number in such a medium. On the
other hand, due to the different dispersion relations a number of $\omega$ mesons cannot
leave the medium\cite{KO} and this obviously increases the probability of a decay inside the
hadronic phase of the fireball.

\begin{figure}[ht]
\centering
\includegraphics[scale=0.35]{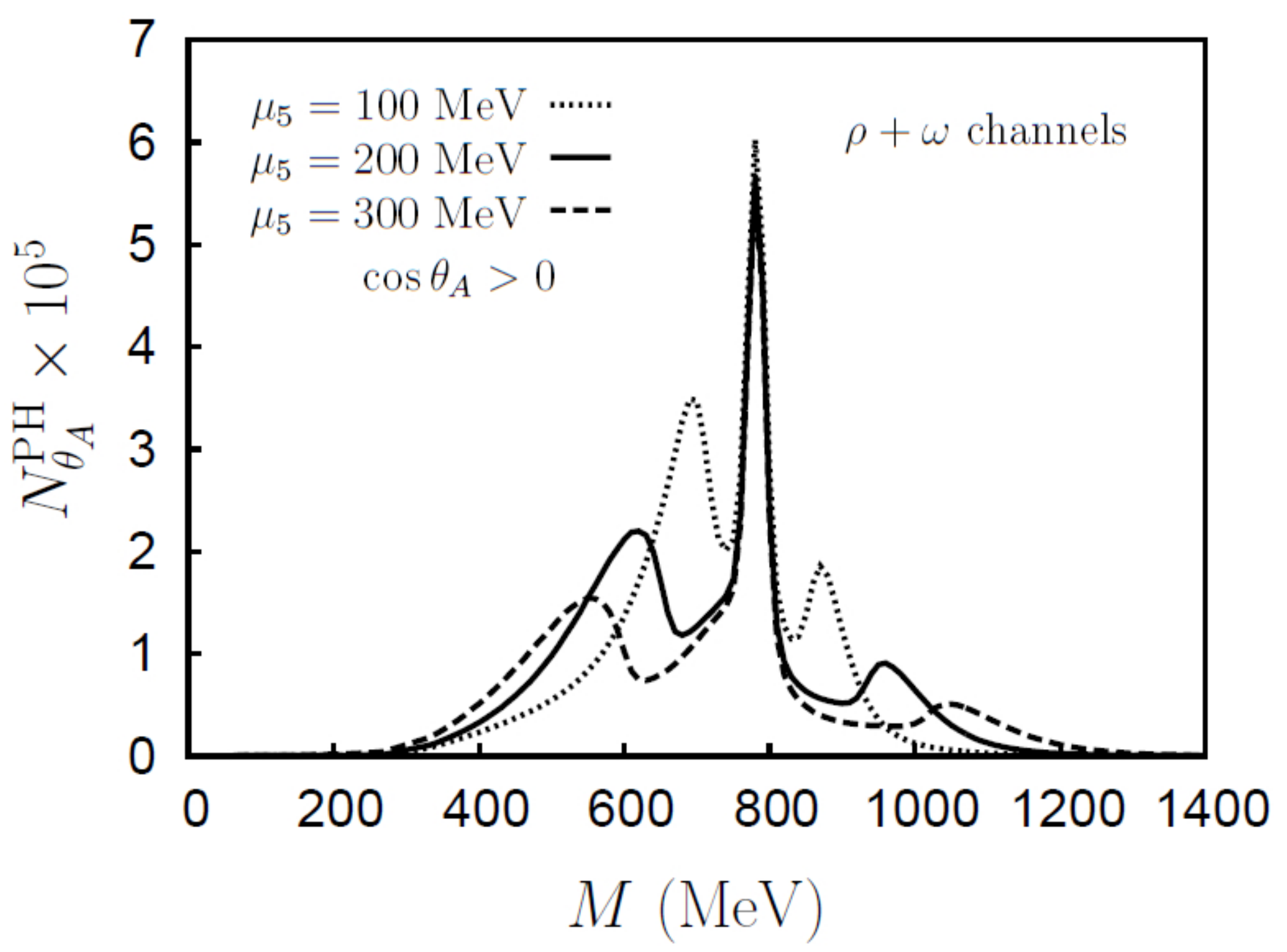}
\caption{Combination of $\rho$ and $\omega$ spectral functions depending on the
invariant mass $M$ and integrating $\cos\theta_A\geq 0$ for $\mu_5=100,200$ and 300 MeV.}
\end{figure}
Now we turn to angle B. The discussion is rather similar. See \cite{angles} for more details.
The main difference with $\theta_A$ is a slightly smaller number of events.
The rest of the analysis is completely equivalent.

\begin{figure}[ht]
\centering
\includegraphics[scale=0.35]{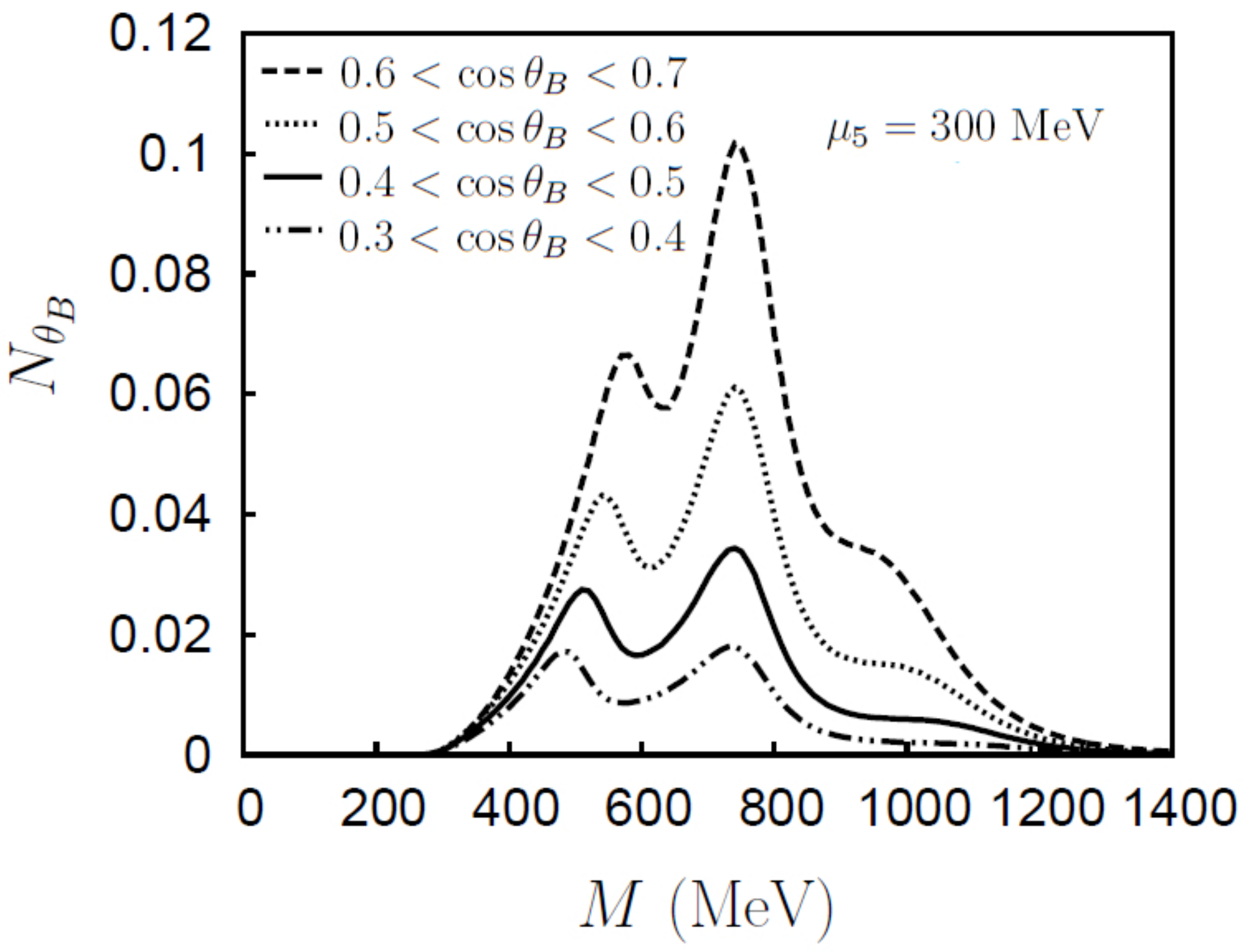} \includegraphics[scale=0.35]{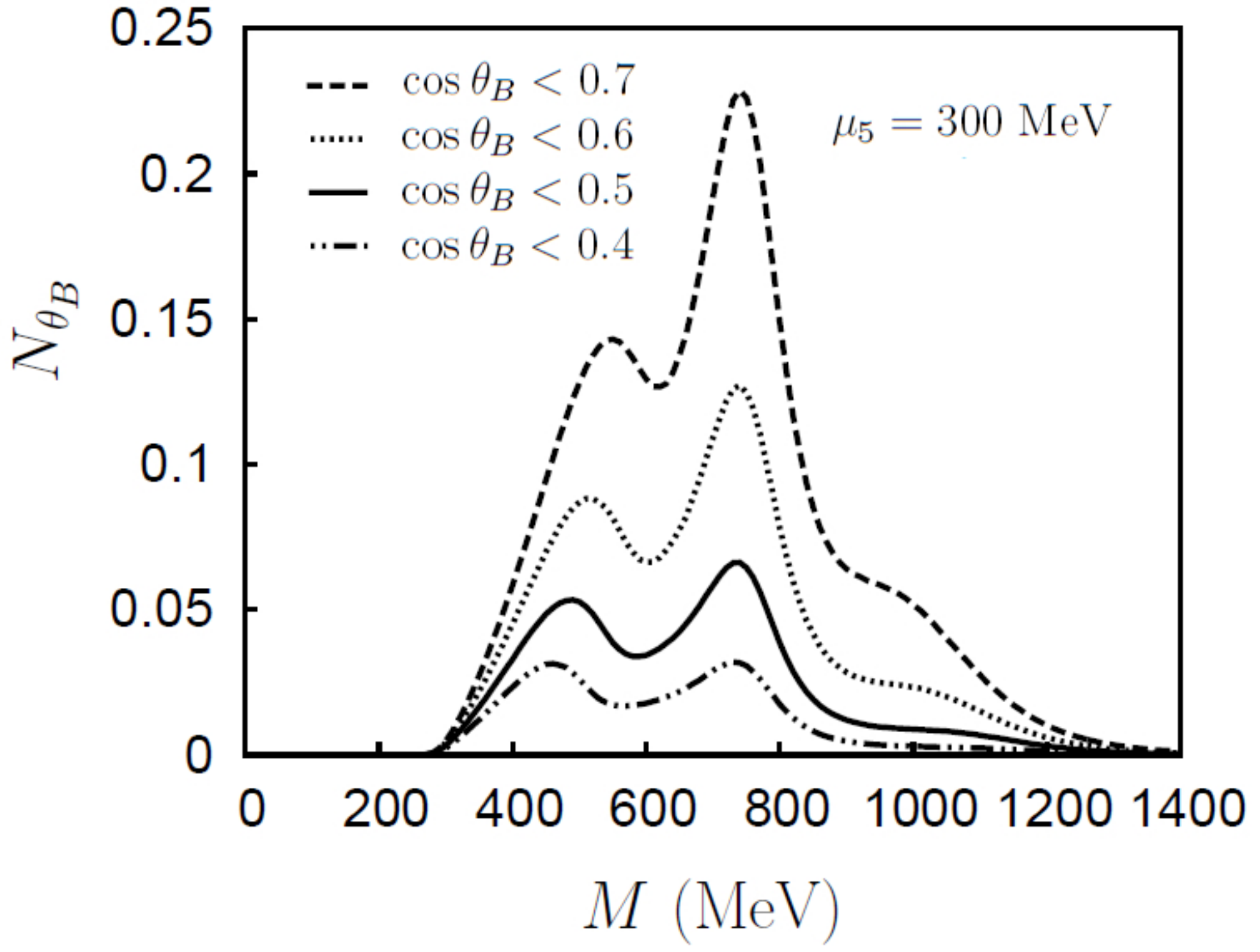}
\caption{Angle $\theta_B$ between one of the two outgoing leptons in the laboratory frame and the same
lepton in the dilepton rest frame. $\rho$ spectral function depending on the invariant mass $M$ for different
ranges of $\theta_B$ for fixed $\mu_5=300$ MeV.}
\end{figure}

\section{Conclusions and outlook}
Let us first list the main results of our work:
\begin{itemize}
\item{$P$- and $CP$-odd effects not forbidden by any physical principle in QCD at finite
density or high temperature, particularly out of equilibrium.}
\item{Topological fluctuations transmit their influence to hadronic physics via an axial
chemical potential for light quarks only.}
\item{$P$- breaking leads to unexpected modifications of the in-medium properties
of scalar and vector mesons.}
\item{Local parity breaking has an influence on the observed dilepton spectrum in
the low-mass region of PHENIX and STAR.}
\item{Some observables may allow us to establish $P$- and $CP$-breaking unambiguously.}
\end{itemize}

Perhaps it is fair to end with some criticisms. To begin with, the possibility that
domains with a non-vanishing chiral charge form and that they grow to a sufficiently large size
is of course unproven. Without this assumption there could be no local parity breaking nor any
of the effects discussed in this presentation. However, we think that it may be possible
to understand this problem (and give an answer) within the glasma picture in the framework
of relativistic hydrodynamics. On the other hand, the situation regarding Dalitz decays
is clearly unsatisfactory. We have reasons to think that they may be substantially
enhanced (and help explain e.g. the STAR data) but it seems very difficult to study
this issue as the effective lagrangian approach as discussed here is not good at all.

All in all, the possibility that signals of parity breaking can be detected in
a strong interaction context in HIC is truly fascinating. But detecting it is
very challenging. We have proposed here a couple
of observables that could possibly yield a signal if local parity breaking is
present.
Experimental collaborations should definitely check this possibility.

\section*{Acknowledgements} It is a pleasure to thank the organizers of the XI Quark
Confinement and Hadron Spectrum Conference for a fruitful meeting and an excellent atmosphere.
This work has been supported through grants FPA2013-46570, 2014-SGR-104 and Consolider
CPAN. A.A. and V.A. were supported by Grant RFBR project 13-02-00127 as
well as by the Saint Petersburg State University Grant.

%%%%%%%%%%%%%%%%%%%%%%%%%%%%%%%%%%%%%%%%%%%%%%%%
%% The bibliography can be prepared using the BibTeX program or
%% manually.
%%
%% The code below assumes that BibTeX is used.  If the bibliography is
%% produced without BibTeX comment out the following lines and see the
%% aipguide.pdf for further information.
%%
%% For your convenience a manually coded example is appended
%% after the \end{document}
%%%%%%%%%%%%%%%%%%%%%%%%%%%%%%%%%%%%%%%%%%%%%%%%

%%%%%%%%%%%%%%%%%%%%%%%%%%%%%%%%%%%%%%%%%%%%%%%%
%% You may have to change the BibTeX style below, depending on your
%% setup or preferences.
%%
%%
%% For The AIP proceedings layouts use either
%%%%%%%%%%%%%%%%%%%%%%%%%%%%%%%%%%%%%%%%%%%%

\bibliographystyle{aipproc}   % if natbib is available
%\bibliographystyle{aipprocl} % if natbib is missing

%%%%%%%%%%%%%%%%%%%%%%%%%%%%%%%%%%%%%%%%%%%
%% You probably want to use your own bibtex database here
%%%%%%%%%%%%%%%%%%%%%%%%%%%%%%%%%%%%%%%%%%%
\bibliography{sample}

%%%%%%%%%%%%%%%%%%%%%%%%%%%%%%%%%%%%%%%%%%%
%% Just a reminder that you may have to run bibtex
%% All of it up to \end{document} can be removed
%% if you don't like the warning.
%%%%%%%%%%%%%%%%%%%%%%%%%%%%%%%%%%%%%%%%%%%
\IfFileExists{\jobname.bbl}{}
 {\typeout{}
  \typeout{******************************************}
  \typeout{** Please run "bibtex \jobname" to optain}
  \typeout{** the bibliography and then re-run LaTeX}
  \typeout{** twice to fix the references!}
  \typeout{******************************************}
  \typeout{}
 }

%%%%%%%%%%%%%%%%%%%%%%%%%%%%%%%%%%%%%%%%%%%
%% The following lines show an example how to produce a bibliography
%% without the help of the BibTeX program. This could be used instead
%% of the above.
%%%%%%%%%%%%%%%%%%%%%%%%%%%%%%%%%%%%%%%%%%%

\end{document}